# PULSE PROPAGATION IN A THREE LEVEL MEDIUM

*G.G. Grigoryan and Y.T. Pashayan*

*Institute for Physical Research, 378410, Ashatarak-2, Armenia*

**Abstract.** We investigate propagation of a pulse pair in a three-level medium under the adiabatic following condition for the general case of unequal oscillator strengths. Exact analytical solutions to the propagation equations have been obtained. It is shown that propagation dynamics strongly depends on the relationship between oscillator strengths. The adiabaticity criterion for the interaction of pulses with three-level media has been derived and analyzed in detail.

Nonlinear interaction of two optical pulses with three-level media under two-photon resonance condition has attracted considerable attention in recent years. It has been studied in the context of electro-magnetically induced transparency (EIT), Stimulated Raman Adiabatic Passage (STIRAP), lasing without inversion (LWI) and other phenomena [1]. All these phenomena rely on the existence of particular superposition states the remarkable property of which is that, when a quantum system is in such a state, there is no interaction between the system and radiation. These states are referred to as "trapped" states and find great application in many fields of laser physics, nonlinear optics, and atomic physics.

While the interaction of a single three-level atom with two optical pulses under trapping conditions is well understood, there is no complete understanding of pulse propagation through a three-level medium. In particular, the trapped state is one of the eigenstates of the instantaneous Hamiltonian and, thus, can be realized only at adiabatic interaction of pulses with an atom and a medium. The adiabaticity criterion and the influence of non-adiabatic corrections have been thoroughly investigated in detail for a single atom in a number of works (see, for example, [2]). However, as known, the interaction adiabaticity, met for a single atom, can break down during propagation into a medium at propagation lengths at which dispersive properties of the medium become essential [3].

From the solutions known the following ones should be mentioned. It is shown in [4] that a pulse pair with arbitrary but identical envelopes ('matched pulses') can propagate in a three-level medium without distortion if the atoms are prepared in coherent superposition of lower states. Another possible mechanism of formstable propagation has been predicted by Eberly ('adiabatons') [5]. Theoretical consideration of the propagation problem with taking into account Doppler broadening has been performed in [6].

In spite of the large number of numerical studies in the field, for example, [7], no analytic solution has been found for the general case of propagation. A review of existing solutions describing some regimes of propagation is presented in [8]. However, all analytical investigations are based on a simplified three-level system with equal oscillator strengths, while actual systems generally involve more complicated situations with unequal oscillator strengths. We show in the present paper that propagation dynamics is greatly affected by the relationship between the oscillator strengths.

The subject of the present paper is the investigation of propagation in a three-level medium with taking into account the first non-adiabatic corrections to the trapped state without restricting the relationship between oscillator strengths. Exact analytic solutions to propagation equations have been found. The analysis of the solutions obtained is given for different parameters of propagation. The propagation in the STIRAP regime is considered thoroughly. It is shown that the evolution of laser pulses differs qualitatively depending on the relationship between oscillator strengths. Simple expressions for the propagation length at which effective population transfer from one quantum state to another is still possible and that at which the pump pulse is completely depleted and reemitted into the Stokes pulse have been given. The adiabaticity criterion for ultrashort pulse propagation in a three-level medium has been obtained.

The paper is organized as follows. In Sec. II we formulate the problem and derive the propagation equations in the adiabatic following approximation [9]. The solutions to the propagation equations are presented in Sec. III. In Sec. IV we focus on the adiabatic behavior of the system and derive the adiabaticity criterion specifying the length at which the interaction adiabaticity breaks down. In Sec. V we study the propagation of counter-intuitive pulse sequence in the STIRAP regime and analyze the evolution of the pulses for different propagation dynamics. The results are summarized in Sec. VI.

**II. Basic formulas**

Three-level systems considered are presented in Fig. 1. States $|1\rangle$, $|2\rangle$ and $|2\rangle$, $|3\rangle$ are connected by the laser radiations $E_p = A_p \cos(k_p x - w_p t + j_p)$ and $E_s = A_s \cos(k_s x - w_s t + j_s)$, respectively. Transitions $|1\rangle \rightarrow |3\rangle$ are dipole-electric forbidden. Pulse durations are assumed to be much less than all relaxation times.

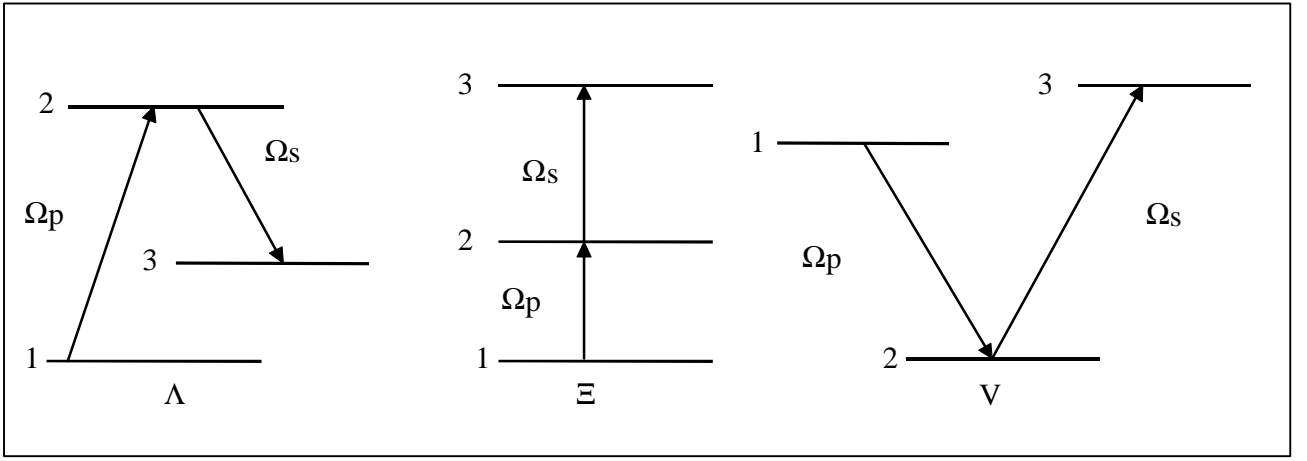

**Fig. 1.** Three-level systems coupled by two resonant pulses with Rabi frequencies $\Omega_p$ and $\Omega_s$.

The population amplitudes of the atomic levels in the resonant approximation satisfy the time-dependent Shroedinger equation

$$i\frac{d}{dt}\begin{pmatrix} b_1 \\ b_2 \\ b_3 \end{pmatrix} = \begin{pmatrix} 0 & W_p & 0 \\ W_p & D_p + i\tilde{A} & W_s \\ 0 & W_s & D_p + D_s \end{pmatrix}\begin{pmatrix} b_1 \\ b_2 \\ b_3 \end{pmatrix}, \quad (1)$$

where $\tilde{A}$ is the decay rate of the upper level; $W_p = -\left|\dfrac{A_p d_p}{\hbar}\right|$, $W_s = -\left|\dfrac{A_s d_s}{\hbar}\right|$ are the Rabi frequencies of the corresponding fields, and $d_p$ and $d_s$ are the dipole moments between levels $|1\rangle$, $|2\rangle$ and $|2\rangle$, $|3\rangle$, respectevely. The detunings off resonance are defined as

$$\begin{array}{ccc} L & X & V \\ D_p = w_{21} - w_p + j_p & D_p = w_{21} - w_p + j_p & D_p = w_{21} + w_p - j_p \\ D_s = w_{32} + w_s - j_s & D_s = w_{32} - w_s + j_s & D_s = w_{32} - w_s + j_s \end{array} \quad (2)$$

The instantaneous Hamiltonian for all three systems has zero eigenvalues under two-photon resonance condition, i.e., when $D_p + D_s = 0$. The corresponding eigenstates are referred to as trapped states and realized under the following initial conditions

$$\frac{b_3(-\infty)}{b_1(-\infty)} = -\frac{W_p(-\infty)}{W_s(-\infty)}, \quad b_2(-\infty) = 0. \quad (3)$$

For example, if an atom is initially in the ground state, i.e., $b_2(-\infty) = b_3(-\infty) = 0$, $b_1(-\infty) = 1$, a counter-intuitive pulse sequence, i.e., the Stokes pulse first and then the pump pulse, should be applied to realize a trapped state.

If $\Delta_p \gg \tilde{A}$, the interaction matrix can be considered real, the Born-Fork procedure is unambiguous [10] and for the atomic state populations in the adiabatic following approximation (with taking into account the first non-adiabatic corrections to the trapped state) we have

$$b_1 = \cos q + i \frac{\dot{q}}{W} \frac{2\sin q}{tg\, 2y}, \quad b_2 = -i\frac{\dot{q}}{W}, \quad b_3 = -\sin q + i \frac{\dot{q}}{W} \frac{2\cos q}{tg\, 2y}, \tag{4}$$

where $W = \sqrt{W_p^2 + W_s^2}$, $tg\, 2y = 2W/D_p$, $tg\, q = W_p/W_s$ ($y \to \frac{p}{4}$ at $D_p \to 0$). When deriving (4) we have made use of the condition $\frac{\dot{q}}{W} \approx (WT)^{-1} \ll 1$ which is the adiabaticity criterion for a single atom at two-photon resonance [1]. The account for the first nonadiabatic correction corresponds to the account for the first-order dispersion of a medium.

The reduced propagation equations in terms of wave variables $x$, $t = t - x/c$ for the $\Lambda$-system in the general case read

$$2W_p \frac{\partial j_p}{\partial x} = -q_p \left(b_1^* b_2 + b_1 b_2^*\right) \qquad 2W_s \frac{\partial j_s}{\partial x} = -q_s \left(b_3^* b_2 + b_3 b_2^*\right)$$

$$2\frac{\partial W_p}{\partial x} = -iq_p \left(b_1^* b_2 - b_1 b_2^*\right) \qquad 2\frac{\partial W_s}{\partial x} = -iq_s \left(b_3^* b_2 - b_3 b_2^*\right), \tag{5}$$

where $q_{p,s} = \frac{2\pi N \omega_{p,s} d_{p,s}^2}{\hbar c}$.

In (5) and all the posterior formulas for a $\Xi$-system $q_s$ should be replaced by $-q_s$ and for a V-system $q_p$ should be replaced by $-q_p$ and $q_s$ by $-q_s$.

It follows from equations (5) that in the ideal adiabatic limit ($b_2=0$) the pulses propagate inside the medium without the shape and phase change (both real and imaginary parts of the dipole moments induced in the medium are equal to zero). However, already the first nonadiabatic correction, accumulating during the propagation process, can change the pattern remarkably. Due to a small non-adiabatic coupling between the trapped state and two other eigenstates the upper level is populated. This leads to non-zero dipole moments between $|1\rangle$, $|2\rangle$ and $|2\rangle$, $|3\rangle$, which, in turn, causes the consequent alteration of both fields.

By introducing the photon number densities $n_{p,s} = \frac{A_{p,s}^2}{\hbar w_{p,s}} = \frac{2pN}{c} \frac{W_{p,s}^2}{q_{p,s}}$ (the dimension of $n_{p,s}$ coincides with that of atomic number density) we have for the total photon number density

$$\frac{\partial(n_p + n_s)}{\partial x} = \frac{2pN}{c}\left(\frac{d}{dt}|b_2|^2 - 2\tilde{A}|b_2|^2\right). \tag{6}$$

For a Ξ-system an analogous relationship is obtained for the difference of the photon number densities

$$\frac{\partial(n_p - n_s)}{\partial x} = \frac{2\pi N}{c}\left(\frac{d}{dt}|b_2|^2 - 2\tilde{A}|b_2|^2\right). \tag{7}$$

Equations (6) and (7) present the conservation law of the total (or the difference) photon number density during propagation under the trapping conditions which is realized in the adiabatic following approximation.

**III. Solutions to propagation equations in the adiabatic following approximation**

In all the posterior formulas we assume one-photon detuning to be large and neglect $\tilde{A}$. Inserting (4) into (5), we have for the Λ-system:

$$\frac{\partial j_{p,s}}{\partial x} = 0, \quad \frac{\partial W_p}{\partial x} = -q_p \frac{\dot{q}\cos q}{W}, \quad \frac{\partial W_s}{\partial x} = q_s \frac{\dot{q}\sin q}{W}. \tag{8}$$

It follows from (8) that phase self-modulation in the medium is absent in this approximation.

Using the definitions of the functions $q(x,t)$ and $W(x,t)$ we have:

$$W_p(x,t) = W(x,t)\sin q(x,t), \quad W_s(x,t) = W(x,t)\cos q(x,t). \tag{9}$$

From the photon number density conservation law we have for the function $W(x,t)$ the following expression:

$$W^2(x,t) = \frac{c}{2pN}\frac{(n_p + n_s)q_p q_s}{q_p|b_1|^2 + q_s|b_3|^2} = W_0^2(t)\frac{q_s \sin^2 q_0(t) + q_p \cos^2 q_0(t)}{q_s \sin^2 q_0(x,t) + q_p \cos^2 q_0(x,t)}, \tag{10}$$

where $W_0^2(t) = W_{p0}^2 + W_{s0}^2$, where $W_{p0}$ and $W_{s0}$ are the functions given at the medium entrance.

Thus, it follows from (4), (9) and (10) that in the adiabatic following approximation the dynamics of the system (both the atomic level population change and the pulse shape evolution) is completely determined by the $q(x,t)$ function which satisfies the equation

$$\frac{\partial q(x,t)}{\partial x} + \frac{q_p \cos^2 q(x,t) + q_s \sin^2 q(x,t)}{W^2(x,t)}\frac{\partial q(x,t)}{\partial t} = 0. \tag{11}$$

Note that in the case of equal oscillator strengths $q_p = q_s$ the $W(x,t)$ function does not change during propagation, $W^2(x,t) = W_0^2(t)$, and equation (11) is simplified substantially.

Introducing "nonlinear group" velocity $u(x,t)$ and using "nonlinear time" $x = t - \frac{x}{u(x,t)}$, for the $q(x,t)$ function we obtain from (11)

$$\frac{\partial q(x,t)}{\partial x} = 0; \quad q(x,t) = q_0(x), \tag{12}$$

where the $x$ function is determined from the implicit equation:

$$\int_\xi^\tau (n_{p0} + n_{s0}) dt' = \frac{x}{q_p q_s} \frac{2\pi N}{c} f^2(\theta(\xi)), \tag{13}$$

where $f(q) = q_s \sin^2 q + q_p \cos^2 q$, and $n_{p0}$ and $n_{s0}$ are the photon number densities at the medium entrance.

Hence, as follows from (4), (9) and (12), two waves, "population wave" and "polarization wave", propagate in the medium with nonlinear group velocity determined by equation (13). In the case of equal oscillator strengths nonlinear propagation velocity is constant if the total photon number density does not depend on time. This propagation regime has been called 'adiabatons'. In the general case for adiabaton formation it is necessary not only photon number density be independent of time but also the quantity $q_s |b_3|^2 + q_p |b_1|^2$ remain constant during propagation process.

**IV. Adiabaticity criterion**

Differentiating (13) on time, we have the following expression for $\dot{q}$

$$\dot{q} = \frac{dq}{dx} \frac{\partial x}{\partial t} = \frac{dq}{dx} \frac{W_0^2(t) f(q_0(t))}{W_0^2(x) f(q_0(x))} \frac{1}{1 - \frac{2x}{W_0^2} \frac{df}{dq_0(x)} \frac{dq_0(x)}{dx}}. \tag{14}$$

As it follows from (14), $\dot{q} \to \infty$ when the denominator in (14) is equal to zero, i.e., when

$$1 - \frac{2(q_p - q_s)x}{W_0^2} \frac{dq_0(x)}{dx} \sin 2q_0(x) \to 0. \tag{15}$$

As $\sin 2q_0(x)$ remains positive over the change range of $q$, relation (15) can only be fulfilled under the following conditions

$$(q_p - q_s) \frac{dq_0(x)}{dx} > 0; \quad \frac{2(q_p - q_s)x}{\Omega_0^2 T} \approx 1. \tag{16}$$

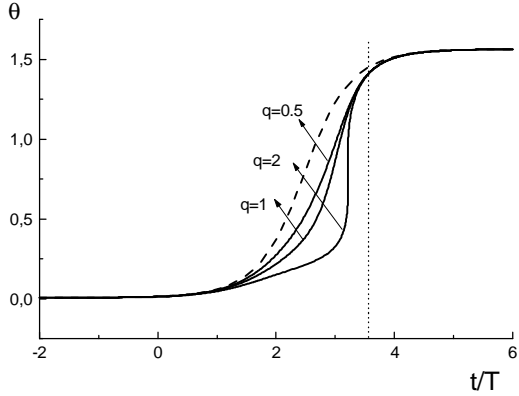

**Fig. 2.** Time evolution of the mixing parameter $q$ for different relationships between $q_p$ and $q_s$ at propagation length $z = 0.03$: $q = 0.5$; 1 and 2. The dashed curve corresponds to the case of $z=0$.

Conditions (16) is just the interaction adiabaticity criterion specifying the critical lengths at which the adiabaticity of pulses, met at the entrance, break down.

Consider this condition in detail. At a counter-intuitive pulse sequence, corresponding to the initial conditions $b_1(-\infty) = 1$, $b_3(-\infty) = 0$, we have $\dot{q} > 0$. In this case the adiabaticity criterion breaks down at $q_p > q_s$. The condition $q_p > q_s$ means that the probability of the transition $|1\rangle \to |2\rangle$ is much more than that of the transition $|2\rangle \to |3\rangle$ and, thus, the population transfer $|1\rangle \to |2\rangle$ dominates the depletion of level $|2\rangle$, i.e., the interaction adiabaticity breaks down. At $q_p \leq q_s$ condition (16) is fulfilled for no values of propagation lengths, and the interaction adabaticity does not break down during the propagation process. Stated above is illustrated by Fig. 2 which presents the time evolution of the mixing parameter $q$ for $q = 0.5$; 1 è 2 (here $q = q_p/q_s$) at the normalized propagation length z = 0.03 ($z \equiv \frac{xq_s}{\Omega_0^2 T}$). The dashed line has been obtained for $z = 0$. The dotted curve corresponds to the case $\dot{q} \to \infty$, i.e., when the adiabatic following condition breaks down. It is seen from the figure that for $q = 2$ the evolution of the parameter differs from the adiabatic one already at this propagation length while in the cases of $q = 0.5$ and 1 the adiabatic evolution of the mixing parameter preserves with propagation.

An intuitive pulse sequence ($\dot{q} < 0$), which can be applied under the initial conditions $b_1(-\infty) = 0$, $b_3(-\infty) = 1$, results in population transfer from level $|3\rangle$ to $|1\rangle$ via $|2\rangle$.

For a $\Xi$-system, replacing $q_s$ by -$q_s$, we have that at a counterintuitive pulse sequence the interaction remains adiabatic at any propagation length, but at an intuitive pulse sequence it can break down at $2x(q_p - q_s)/W_0^2 T \approx 1$. For a V-system replacing $q_s$ by -$q_s$ and $q_p$ by -$q_p$, we have the case just opposite to a $\Lambda$-system.

Thus, the interaction adiabaticity is rather sensitive to the relationship between oscillator strengths. It preserves when $q \leq 1$ and breaks down rather quickly with the propagation length when $q > 1$. It should be noted, that this conclusion contradicts with that of [11] whose authors state that the interaction adiabaticity breaks down for any relationship between $q_p$ and $q_s$. Note that $q$ can be written in the form $q = \dfrac{n_s - 1}{n_p + 1}$, where $n_s$ and $n_p$ are the refraction coefficients for the Stokes and pump pulses, respectively.

### V. Propagation in the STIRAP regime

Using the solutions obtained we have investigated the time evolution of the normalized Rabi frequencies of counter-intuitive pulse sequence in a $\Lambda$-system at different propagation lengths for different relationships between the oscillator strengths ($q \leq 1$).

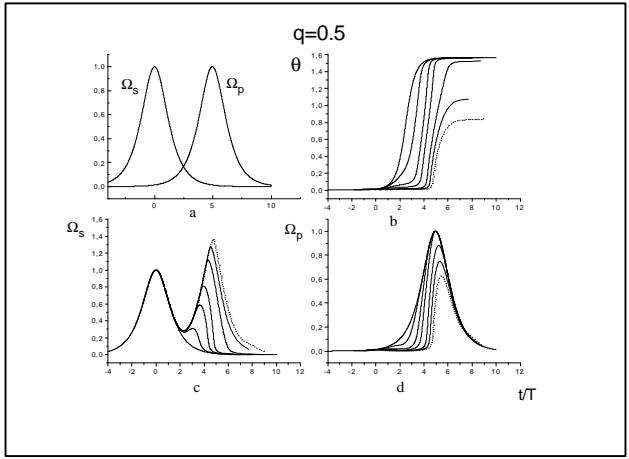
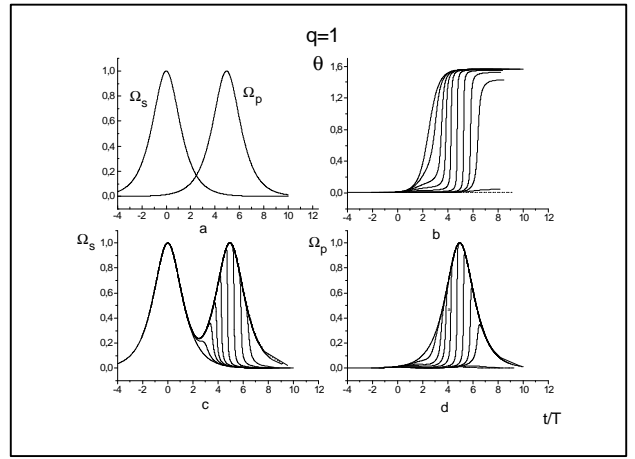

**Fig. 3.** Time evolution of
a) the Rabi frequencies of the pump and Stokes pulses at the medium entrance; b) the mixing parameter $\theta$; c) the Rabi frequencies of the pump pulse for different propagation lengths ($z$ changes from 0 to 3.3); d) the Rabi frequencies of the Stokes pulse for different propagation lengths ($z$ changes from 0 to 3.3). The relationship between $q_p$ and $q_s$ is equal to 0.5. Different curves correspond to different values of the propagation length.

**Fig. 4.** Time evolution of
a) the Rabi frequencies of the pump and Stokes pulses at the medium entrance; b) the mixing parameter $q$; c) the Rabi frequencies of the pump pulse for different propagation lengths ($z$ changes from 0 to 3.3); d) the Rabi frequencies of the Stokes pulse for different propagation lengths ($z$ changes from 0 to 3.3). The relationship between $q_p$ and $q_s$ is equal to 1. Different curves correspond to different values of the propagation length.

In Figs. 3-4 we present the time evolution of the normalized Rabi frequencies of both pulses as they propagate inside the medium for $q = 0.5$ and $q = 1$, respectively. We consider the hyperbolic secant pulse shapes: $\Omega_p(0,t) = \dfrac{a}{ch(t - t_d)/T}$, $\Omega_s(0,t) = \dfrac{a}{cht/T}$, where $t_d$ is the delay time between the pulses.

It is seen that both pulses experience significant reshaping as they propagate into the medium. The leading edge of the pump pulse undergoes gradual depletion. At the tailing edge of the Stokes pulse there appears an additional peak the amplitude of which increases continuously during the propagation. Thus, there occurs a significant dynamical redistribution of the photon number in the overlapping range. Comparisons clearly show that the redistribution dynamics depends on the value of $q$. Indeed, for the case where $q = 0.5$ at the propagation length of $z = 3$ (the dotted curve in the figures) the pump pulse is still intense enough while at the same length for the other case it is completely depleted and reemitted into the Stokes pulse. Thus, the regime of effective photon transfer from the pump pulse into the Stokes one is possible during propagation. The complete pump depletion occurs when $\sin^2 q_0(x) = 0$, i.e. $x \leq -T + t_d$. The following simple expression, specifying the critical length at which a complete depletion of the pump pulse occurs, has been obtained from (13)

$$z_{pump} \sim \frac{2+q}{q^2}. \tag{17}$$

So, the more $q$ is, the quicker the regime is set. In particular, for the case $q = 1$ this length is equal to $z \sim 3$, which agrees with Fig. 4 (the dotted curve). Let now estimate the value of $z$. Estimations show that $z=1$ corresponds to $\sim 48$ cm. Thus, for $q = 1$ at the propagation length of 144 cm the pump pulse is completely reemitted into the Stokes one.

Figs. 3-4 also present the time evolution of the mixing angle $q(x,t)$. It is seen that initially effective STIRAP from the ground state to the final state is possible. However, as the pulses propagate into the medium, the efficiency of the transfer decreases. It is explained by the following. It is known that for effective STIRAP not only a counter-intuitive pulse switching on ($q(-\infty) \to 0$) but also a corresponding switching off ($q(+\infty) \to p/2$) is required. For such a switching off it is necessary that at $t \to +\infty$ $W_p \neq 0$ but $W_s \to 0$. It is seen from the figures that at the initial propagation lengths this condition is fulfilled and the end value of $q$ is close to $p/2$. However, after a certain propagation length there occurs a noticeable decrease in the amplitude of the pump pulse and the corresponding amplification of the Stokes pulse at its tail. Thus, after this length the conditions $W_p(+\infty) \neq 0$, $W_s(+\infty) \to 0$, required for effective STIRAP, break down and the end value of $q$ is different from $p/2$, which means that the transfer process is not complete.

Consequently, we see that even if all the conditions for effective STIRAP are met at the entry surface of the medium, during propagation they may break down due to reshaping effects.

Find now the expression specifying the propagation length at which effective STIRAP is still possible. After the interaction with the pulses the atoms will be in the final state if $\sin^2\theta_0(\xi)=1$ and $\cos^2 q_0(x)=0$, which corresponds to $\xi \geq T$. Then for (13) we have

$$\int_T^{T+t_d}(n_{p0}+n_{s0})dt' = \frac{x}{q_p q_s}\frac{2pN}{c}q_s^2 \qquad (18)$$

or $$\bar{n}_0 t_d = \frac{q_s}{q_p}\frac{2pN}{c}x, \qquad (19)$$

where $\bar{n}_0$ is the number of photons in the pump pulse after the Stokes pulse switching off. Substituting (19) into (18) one can easily obtain the following expression for the propagation length at which STIRAP is still possible

$$z_{stirap} \sim t_d/2T. \qquad (20)$$

It follows from (19) and the inset that for the given time delay $t_d$ between the pulses the more the number of photons in the time interval between $T$ and $T+t_d$, the more this length. This means that the Stokes pulse should be switched off sharply, while the pump pulse should be switched off more smoothly. Thus, by corresponding pulse switching off, namely, more smooth for the pump pulse and more sharp for the Stokes one, one can provide more photons in the pump pulse and so the longer the penetrating length for effective STIRAP. So, effective STIRAP takes place at asymmetric switching off of the pump and the Stokes pulses.

On the other hand, as follows from (20), the more $t_d$, the more the effective STIRAP process and the longer the STIRAP penetration length. It should be noted that $z_{stirap}$ should not exceed $z_{pump}$ as at this length the pump pulse, actually, is completely depleted and the STIRAP process is not possible.

Thus, the propagation length at which effective STIRAP is possible depends only on the time delay between the pulses but not on the relationship between $q_p$ and $q_s$. This conclusion is confirmed by Fig. 5. This figure presents the time evolution of the ground state population at different time delays between the pulses for $q = 0.5$ and $q = 1$. Comparisons show that the penetrating length of STIRAP increases with the increase of $t_d$. Indeed, for $t_d = 2.5$ and $z = 1.6$ the efficiency of the process decreases while at the same propagation length at $t_d = 5$ the process is intense enough, i.e. according to (20) the more $t_d$, the more effective the STIRAP process. It is seen from the figure that the final values of the ground state population do not differ for $q = 0.5$ and $q = 1$, i.e., $z_{stirap}$ does not depend on the relationship between the oscillator strengths. Some difference between the population dynamics is explained by the fact that the time evolution of the pulses depends on the value of $q$.

**Fig. 5.** Population transfer for $q = 0.5$ and $q = 1$. Shown are the populations of the initial and final states for $t_d = 2.5$ and $t_d = 5$. Different curves correspond to different propagation length.

Fig. 6 presents the case where $q = 0.001$. It is seen from the figure that the pump pulse propagates without changing its shape while the Stokes pulse changes significantly, which is not surprising as the photon number preserves during propagation, which is confirmed by Fig. 6c. The figure presents the change of the photon number as the pulses propagate into the medium. Some amplification of the pump pulse, observed at the tailing edge, indicates the beginning of the reverse process of the photon transfer from the Stokes pulse into the pump pulse. This is the beginning of the process of adiabaton formation. However, in the case considered adiabatons can not be formed as the Stokes pulse switches off and the process stops.

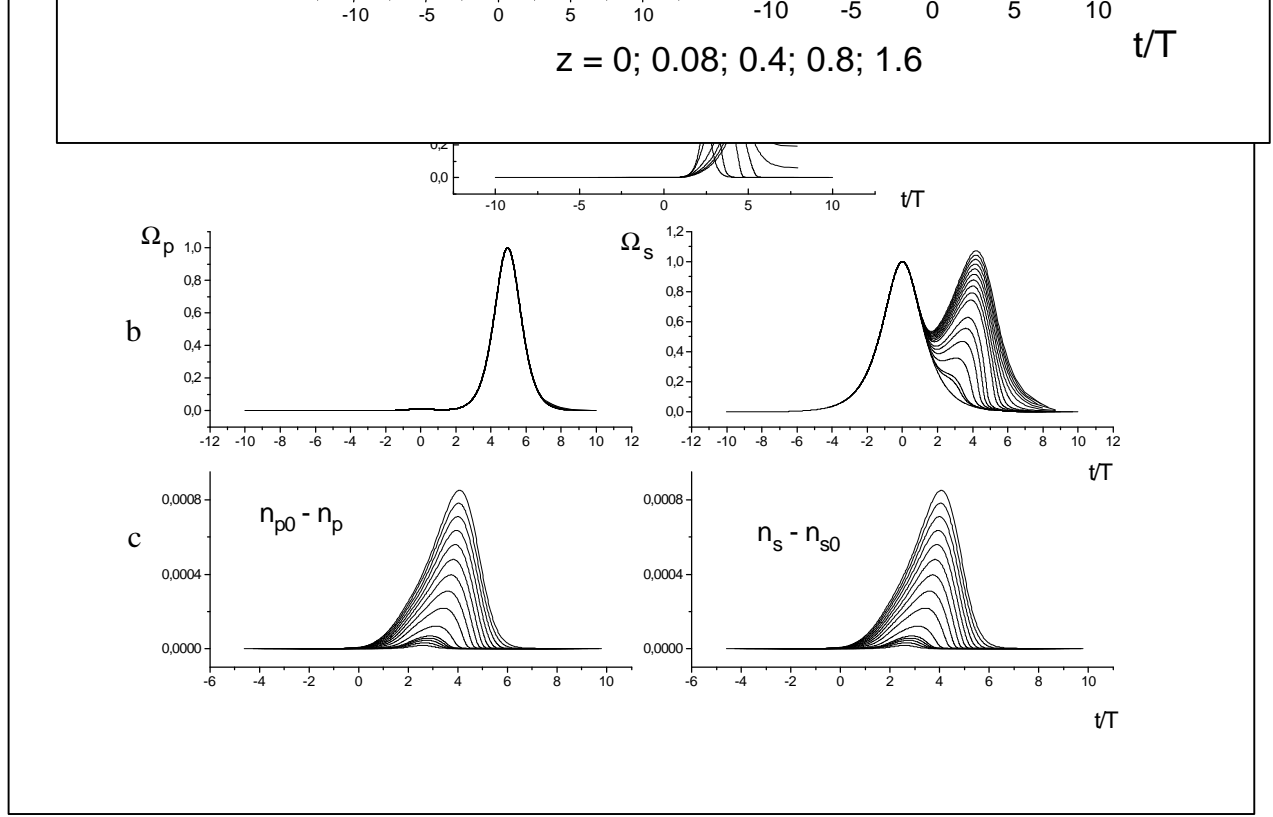

**Fig. 6.** Time evolution of
a) the initial and final state population; b) the Rabi frequencies of the pump and Stokes pulse;
c) the change of the photon number in the pump and Stokes pulses. The relationship between $q_p$ and $q_s$ is equal to 0.001. Different curves correspond to different values of the propagation length.

## VI. Summary.

In the present paper we have studied the propagation of partially overlapping pulses through a three-level medium of $\Lambda$, $\Xi$ and $V$ types under the adiabatic following condition. We have obtained exact analytical solutions to the propagation equations for the general case of unequal oscillator strengths and show the importance of this parameter for propagation process. The investigation performed shows that propagation dynamics is strongly affected by the relationship between oscillator strengths.

We derived the adiabaticity criterion for the radiation interaction with a three-level medium. We show that the interaction adiabaticity strongly depends on the relationship between oscillator strengths $q$. In the case of a counter-intuitive pulse sequence the interaction adiabaticity, met at the medium entrance, preserves for any value of propagation lengths if $q \leq 1$ and breaks down rather quickly when $q > 1$.

Next we study the evolution of pulses during propagation under the adiabatic following condition for the propagation parameters for which interaction adiabaticity is met. The pulses have been found to exhibit different behavior depending on the relation between oscillator strengths. The analysis performed shows that during propagation under the adiabatic following condition there occurs a considerable pulse reshaping which lies in gradual depletion of the leading edge of the pump pulse and corresponding amplification of the tailing edge of the Stokes pulse. Such pulse shape change during propagation leads to decreasing the efficiency of STIRAP.

From the analytical solutions obtained simple expressions for the propagation length at which a complete reemitting of the pump pulse into the Stokes pulse and that at which STIRAP is possible have also been obtained.

**ACKNOWLEDJEMENTS.**

The authors wish to thank Dr. R. Unanyan for stimulating discussions. The work was supported by the State sources of the Republic of Armenia under scientific theme 96-772.